# A Floating Point Division Unit based on Taylor-series expansion algorithm and Iterative Logarithmic multiplier

Riyansh K. Karani, Akash K. Rana, Dhruv H. Reshamwala, Kishore Saldanha

Electronics Department, Dwarkadas J. Sanghvi College of Engineering, Mumbai
riyansh.karani.011235@gmail.com
akash9182akash@gmail.com
dhr.reshamwala@gmail.com
kishoresaldanha@gmail.com

## ABSTRACT

*Floating point division, even though being an infrequent operation in the traditional sense, is indispensable when it comes to a range of non-traditional applications such as K-Means Clustering and QR Decomposition just to name a few. In such applications, hardware support for floating point division would boost the performance of the entire system. In this paper, we present a novel architecture for a floating point division unit based on the Taylor-series expansion algorithm. We show that the Iterative Logarithmic Multiplier is very well suited to be used as a part of this architecture. We propose an implementation of the powering unit that can calculate an odd power and an even power of a number simultaneously, meanwhile having little hardware overhead when compared to the Iterative Logarithmic Multiplier.*

## 1. INTRODUCTION

Approximation methods such as Newton-Raphson and Taylor-series can be used to approximate functions where direct computation of these functions is either computationally very expensive or not possible. The Taylor-series expansion is an approximation method that generates a high order polynomial approximation of a function at some value in its domain. The idea is simple; given that a function and its first $n$ derivatives are continuous at some point in its domain, the function can be approximated by a polynomial of degree $n$ at that point. The higher the order of this polynomial, the better is the approximation. Following on this idea, the reciprocal of a number can be approximated as a very simple Taylor-series polynomial, and thus the problem of division of one number by another is essentially reduced to multiplication of one number and the Taylor-series polynomial of the other [6]. Calculating the terms of the Taylor-series polynomial that approximates the reciprocal of a number $x$ involves calculating the powers of $x$ itself $(x^2, x^3, \cdots)$, where calculating each higher power generates a better approximation.

Naturally, there is a need of a multiplier unit to calculate these powers, and the performance of the floating point division unit then depends almost solely on the performance of the multiplier unit. There are many popular multiplier architectures in use today [3]. The Iterative Logarithmic Multiplier is one such multiplier architecture [12]. What makes it an attractive choice is that it is highly programmable. The accuracy of the product generated by this multiplier can be precisely controlled, which comes in very handy when one does not need full precision multiplication, like in the case of digital signal processing. But perhaps what is even more important is that because of its inherent nature, its implementation is very hardware efficient when it comes to computation of squares. This is because the mathematical description of the Iterative Logarithmic Multiplier is quite simplified when multiplying a number with itself, as when compared to multiplying two different numbers. Because every even power of a number $(x^{2k})$ can be represented as a square of a smaller power of the number $(x^k)$, every alternate power of $x$ is representable as a square of some other number. Hence, the Iterative Logarithmic Multiplier is a very suitable candidate when choosing a multiplier architecture for calculating the terms of the Taylor-series polynomial approximation of a reciprocal.

In this paper, we present an architecture for a floating point division unit based on the Iterative Logarithmic Multiplication algorithm, and the Taylor-series expansion algorithm. The proposed architecture is designed to be hardware efficient, as is the requirement when designing architectures for high speed computational units. We start by describing the Taylor-series expansion algorithm in section II, and analyse the approximation errors generated when using this approach. We then introduce the methodology for calculating the reciprocal of a number using the Taylor-series approach, and derive a quantitative measure of the generated error. Since the approach above requires an initial approximation of the reciprocal, in section III, we start by describing and analysing linear approximation as a possible approach. We then build on this to present the piecewise linear approximation based approach that we have employed in our implementation of the floating point division unit. In section IV, we describe the Iterative Logarithmic Multiplier as proposed by Babić, Avramović and Bulić [12], followed by the discussion and implementation of the proposed squaring unit in section V. Finally, we present the architecture for the proposed powering unit in section VI, and discuss its implementation and features.

## 2. Taylor-series Expansion Algorithm

A Taylor-series is a series expansion of a function at a point in its domain. Let $f(x)$ be a function such that its first $n$ derivatives are continuous. Then, $f(x)$ at a point $x = a$ can be approximated as a Taylor-series expansion as follows [1]

$$f(x) \approx \sum_{k=0}^{n} \frac{f^{(k)}(a)}{k!}(x-a)^k \qquad (1)$$

where, $f^{(k)}$ is the $k^{th}$ derivative of $f$. Because equation (1) gives the approximate value of $f(x)$ at $x$ near $a$, it is necessary to estimate the error as a function of $x$. Another formulation of the Taylor-series, called the *Taylor Series with Remainder* [1] [2] is given as

$$f(x) = \sum_{k=0}^{n} \frac{f^{(k)}(a)}{k!}(x-a)^k + E_n(x) \qquad (2)$$

where $E_n(x)$ is the error term, and is given as

$$E_n(x) = \frac{1}{n!} \int_a^x (x-t)^n f^{n+1}(t) dt \qquad (3)$$

Assume that the values of $x$ are bound to the close interval $[a-c, a+c]$. Then, using the Mean Value Theorem, it can be proved [1] that there exists a value $x = \xi$ such that

$$\int_a^x (x-t)^n f^{n+1}(t) dt = f^{n+1}(\xi) \int_a^x (x-t)^n dt$$

Solving the integral, we get

$$\int_a^x (x-t)^n f^{n+1}(t) dt = \frac{f^{n+1}(\xi)}{(n+1)}(x-a)^{n+1} \qquad (4)$$

for some $\xi \in [a-c, a+c]$. Then, the error term becomes

$$E_n(x) = \frac{f^{n+1}(\xi)}{(n+1)!}(x-a)^{n+1} \tag{5}$$

Although this formulation of the error term does not precisely determine its value, it lets us determine the bounds on the size of the error term. Since $f^{n+1}$ is continuous in $[a-c, a+c]$,

$$There\ exists\ M\ \forall\ \xi\ \in\ [a-c, a+c],\ f^{n+1}(\xi) \leq M \tag{6}$$

Hence,

$$0 \leq |E_n(x)| \leq \frac{|M|}{(n+1)!}|(x-a)|^{n+1} \tag{7}$$

which means that

$$|E_n(x)| = o((x-a)^n) \tag{8}$$

Thus, we obtain an upper bound on the value of $E^n(x)$, and we can say that if the value of $x$ is close enough to that of $a$, then as $n$ increases, the error becomes smaller.

The Taylor-series for $f(x) = \dfrac{1}{1-x}$ can be written as

$$f(x) = \frac{1}{1-x} \approx \sum_{n=0}^{\infty} x^n\ about\ x = 0 \tag{9}$$

Suppose we wish to compute the value of $x^{-1}$. Let $y_0$ be approximately equal to $x^{-1}$. Then, $y_0 x \approx 1$. From equation (9), we can write

$$\frac{1}{1-(1-xy_0)} \approx 1 + (1-xy_0) + (1-xy_0)^2 + (1-xy_0)^3 + \cdots \tag{10}$$

And thus,

$$\frac{1}{x} \approx y_0 \left( \sum_{k=0}^{\infty} (1-xy_0)^k \right) \tag{11}$$

The error term can be calculated from (5) as

$$E_n(x, y_0) = \frac{(1-xy_0)^{n+1}}{(1-\xi)^{n+2}} \tag{12}$$

where $\xi$ is some value of $(1-xy_0)$.

What equation (11) means is that starting from an approximate value of $x^{-1}$, we can calculate an arbitrarily accurate value for it, and this precision depends on the highest power $k$ of the Taylor-series polynomial.

## 3. Initial Approximation

As stated in the previous section, we need an initial approximation of the inverse of a number, in order to calculate a more precise approximation using the Taylor-series expansion algorithm, and according to equation (7), the number of iterations required to obtain an approximation with a desired precision depends on the initial approximation. So, it is very important to select an appropriate method for finding an initial approximation. There are different kinds of methods [5] based on linear approximation, direct lookup tables [7] [8] [11], table lookup followed by multiplication [4] and polynomial approximations [9]. In our implementation, we choose a different approach. We employ a piecewise linear approximation for generating the initial approximation, and we show that one can obtain any desired amount of precision using this method, without much increase in complexity.

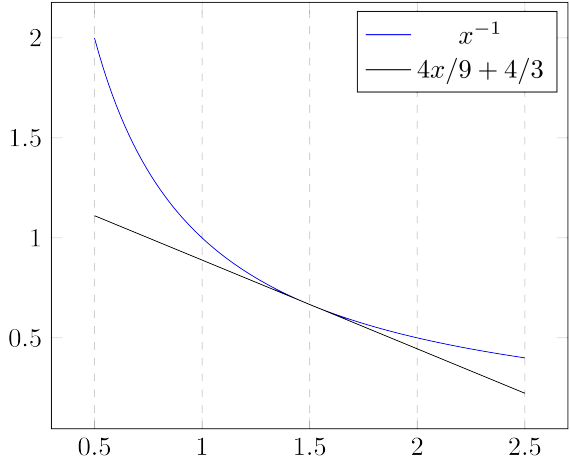
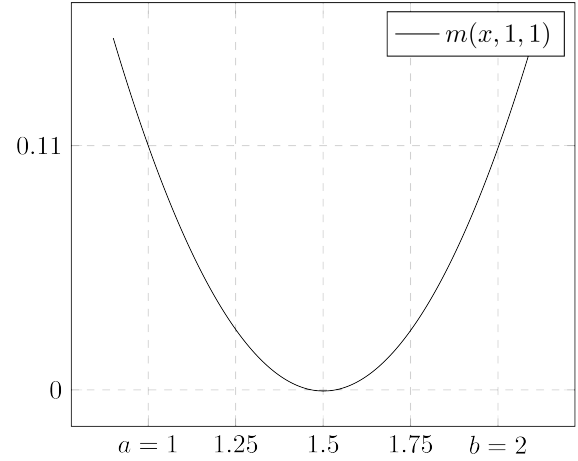

Figure 1: Plot showing the actual and linear approximation values of $x^{-1}$, for $a = 1$ and $b = 2$

Figure 2: Plot showing the values of $m$ for $x$ in the range $[1, 2]$

Initially, consider a linear approximation to the reciprocal of $x$ as shown in Figure 1. The error in the approximation at any value of $x$ can be written as

$$E = \frac{1}{x} + \frac{x}{p^2} - \frac{2}{p} \tag{13}$$

Suppose we are interested in calculating the reciprocals of numbers in the range $[a, b]$, the total error (for all the possible numbers in this range) can be expressed as

$$E_{total} = \int_a^b E \, dx = log_e(b/a) + \frac{(b^2 - a^2)}{2p^2} - \frac{2(b-a)}{p} \tag{14}$$

Since $p$ is a variable, we can find the value of $p$ such that $E_{total}$ is minimum by differentiating (14) w.r.t $p$ and equating it to zero. Doing so, we find that $E_{total}$ is minimum for $p = (a+b)/2$. The optimum linear approximation of $x^{-1}$ is then

$$y_0 = \frac{-4\,x}{(a+b)^2} + \frac{4}{a+b} \tag{15}$$

Let

$$m(x, a, b) = 1 - x \cdot y_0 \tag{16}$$

Observe from equation (10) that the term $1 - xy_0$ is the same as $m$ in equation (16). Equation (12) then gives us the error in the approximation of $x^{-1}$. Figure 2 shows a plot of $m$ vs $x$ for $x$ in range $[a, b]$. since the error term in equation (12) is directly proportional to $\xi$, the error is maximum when $x$ is either $a$ or $b$. Hence

$$E_n(x, y_0) \leq \left(\frac{(a+b)^2}{4ab}\right)^{n+2} (1 - xy_0)^{n+1} \tag{17}$$

From equation (17), we can precisely calculate the minimum number of iterations needed to calculate the reciprocal of a number $x$ up to a required precision, given an initial approximation $y_0$. Assuming that the number $x$ is the significand of a floating point number represented in the IEEE-754 format (which is generally the use case of a floating point division unit), $x$ is normalized and thus $a$ is 1 and $b$ is 2. Hence, in this case

$$E_n(x, y_0) \leq \left(\frac{9}{8}\right)^{n+2} (1 - xy_0)^{n+1} \qquad (18)$$

Assuming the worst case ($x = 1$, where the initial error is maximum), one can calculate that to obtain at least 53 bits of precision (which is the maximum precision required for a 64-bit floating point number), we need a maximum of 17 iterations using this approach. We can reduce this number by employing a piecewise linear approximation of the reciprocal, instead of just a linear approximation. Initially, assume that the total range of $x$ is divided into two segments of the same length. Using equation (14) for calculating the total error for the two segments, we see that $E_{total}$ for the first segment is greater that that for the second segment. According to equation (7), we need to account for the maximum error while calculating the maximum number of iterations required to achieve a particular precision. Hence, the maximum number of iterations required to compute the reciprocal will still be bounded above by the maximum error from the first segment. Thus, it would make more sense to sacrifice some of the accuracy in the second segment in order to improve the accuracy in the first segment, by reducing the value of $p$. The most optimum solution would then be in the case when the total error in both the segments is the same. By equating the values of $y_0$ at $x = p$, we find out that $E_{total}$ for both the segments is the same when $p = \sqrt{ab}$. Using this approach for 64-bit floating point numbers, equation (17) tells us that a minimum of 15 iterations is required for calculating the reciprocal of the significand up to a precision of 53 bits. This is only a little improvement over the previous number, but we can extend this concept to more than two segments, in order to achieve the accuracy we want.

The following is the general procedure that can be followed in order to calculate the number of segments, and the location of each segment required for the piecewise linear approximation of a 64-bit IEEE-754 floating point number, starting from a known number of iterations $n$:

1) Select a value for the number of iterations $n$, and the maximum precision $pr_{max}$ of the computed reciprocal. Since the number of iterations required to compute the reciprocal is maximum for $x = 1$, we start with $a = 1$ and let $b_0$ be the end of segment 1. Then, according to equation (17)

   $$\left(\frac{(1+b_0)^2}{4b_0}\right)^{n+2} \cdot \left(m(1,1,b_0)\right)^{n+1} \leq \frac{1}{2^{pr_{max}}}$$

   Substituting the value for $m$

   $$\frac{(1+b_0)^2}{(4b_0)^{n+2}} \cdot (1-b_0)^{2n+2} \leq 2^{-53} \qquad (19)$$

   We can thus calculate the maximum value of $b_0$ for the chosen value of $n$ that can satisfy the above relationship.

2) Once we have the value of $b_0$, by the same logic as the one we employed in the two segment case, we consider the point $(b_0, y_0(b_0))$ as the starting point for the next segment and repeat the above procedure for $x = a = b_0$ to find the value of $b_1$. More generally

   $$\frac{(b_{k-1}+b_k)^2}{(4 \cdot b_{k-1} \cdot b_k)^{n+2}} \cdot (b_{k-1}-b_k)^{2n+2} \leq 2^{-53} \qquad (20)$$

3) Repeat the above procedure to obtain the values for $b_2, b_3, \cdots, b_k$ until $b_k \geq 2$, as shown in Figure 3.

Table I shows the values for $b_k$ for $n = 5$ that are derived using equation (20). Using only 8 segments for the piecewise linear approximation, we can bring down the initial approximation to such a small value that after a maximum of 5 iterations, we get a precision of at least 53 bits.

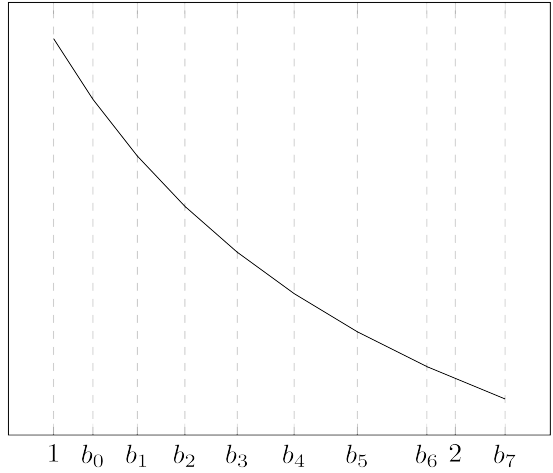

Figure 3: Piecewise linear approximation of $x^{-1}$ for $x$ in the range $[1, 2]$, derived for $n = 5$

Table I: Piecewise Linear approximation segments

| | |
|---|---|
| $a$ | 1 |
| $b_0$ | 1.09811 |
| $b_1$ | 1.20835 |
| $b_2$ | 1.3269 |
| $b_3$ | 1.45709 |
| $b_4$ | 1.59866 |
| $b_5$ | 1.75616 |
| $b_6$ | 1.92922 |
| $b_7$ | 2.12392 |

## 4. Iterative Logarithmic Multiplier

Logarithmic Number System (LNS) based multipliers are a good choice when there is a possibility of trading accuracy for speed (such as in Digital Signal Processing). The main advantage of LNS based multipliers is the substitution of multiplication with addition, which is a much simpler operation in terms of complexity. LNS multipliers can be divided into two categories [12], one based on methods that use lookup-tables, and the others based on Mitchell's algorithm [10]. The major drawback with Mitchell's algorithm is the error in the product due to the piecewise linear approximation of the logarithmic curve. The Iterative Logarithmic Multiplier, as the name suggests, proposes an iterative solution to computer this error term, and hence generate a better approximation to the product.

The binary representation of a number can be written as

$$N = 2^k(1 + \sum_{i=0}^{k-1} 2^{i-k} B_i) = 2^k(1 + x) \qquad (21)$$

where, $B_i$ is the $i^{th}$ bit of $N$, $x$ is the mantissa, and $k$ is the total number of bits in $N$.

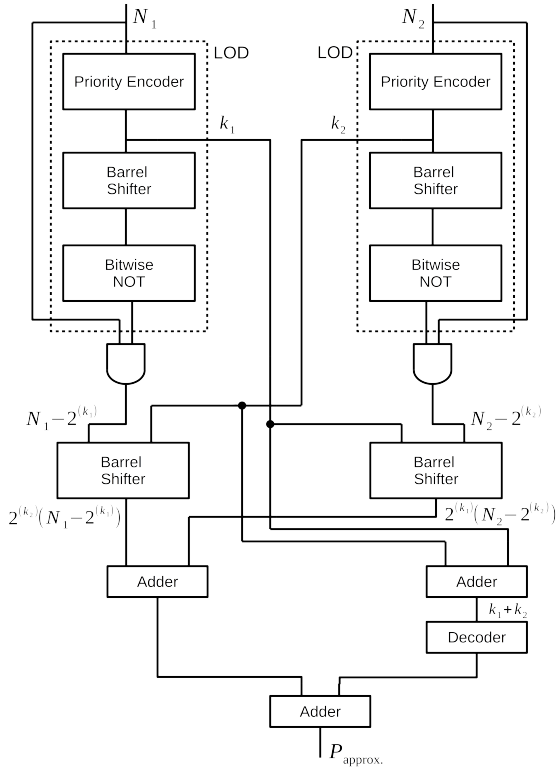 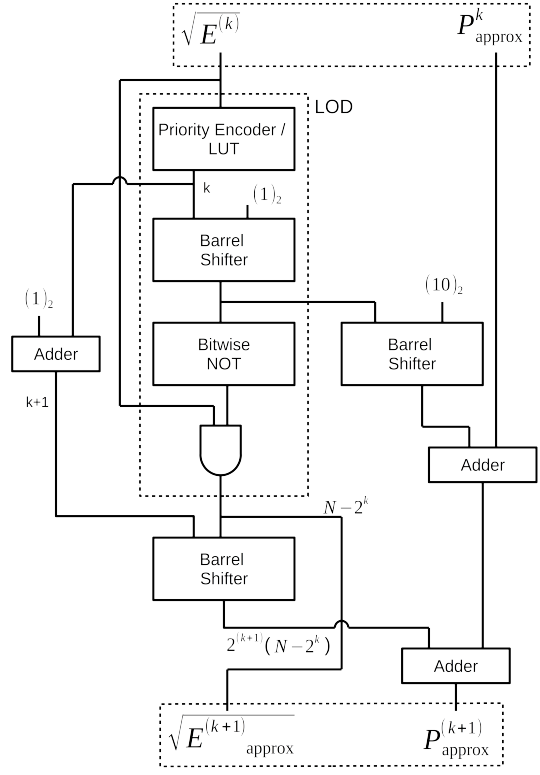

Figure 4: Block diagram of an Iterative Logarithmic Multiplier

Figure 5: Architecture of the proposed squaring unit

Employing this representation, the product of two numbers $N_1$ and $N_2$ can be written as
$$N_1 \cdot N_2 = 2^{k_1+k_2}\left(1 + x_1 + x_2\right) + 2^{k_1+k_2}\left(x_1 \cdot x_2\right) \qquad (22)$$
From equation (21)
$$x \cdot k = N - 2^k$$
Thus, equation (22) can be rewritten as
$$N_1 \cdot N_2 = 2^{k_1+k_2} + 2^{k_2}(N_1 - 2^{k_1}) + 2^{k_1}(N_2 - 2^{k_2}) + (N_1 - 2^{k_1})(N_2 - 2^{k_2}) \quad (23)$$
The error in Mitchell's algorithm is because of ignoring the second term in equation (22). Let
$$P_{approx}^{(0)} = 2^{k_1+k_2} + 2^{k_2}(N_1 - 2^{k_1}) + 2^{k_1}(N_2 - 2^{k_2}) \qquad (24)$$
$$E^{(0)} = (N_1 - 2^{k_1})(N_2 - 2^{k_2}) \qquad (25)$$
Then, the product can be written as
$$N_1 \cdot N_2 = P_{approx}^{(0)} + E^{(0)} \qquad (26)$$
Observe that $(N_1 - 2^{k_1})$ is nothing but $N_1$ with its $k_1^{st}$ bit cleared, and thus the computation of $E^{(0)}$ can be reduced to multiplication of two different numbers. Following this logic, we can write
$$E^{(0)} = P_{approx}^{(1)} + E^{(1)}$$
$$E^{(1)} = P_{approx}^{(2)} + E^{(2)} \qquad (27)$$
$$\ldots$$

By iterating over this process until one of the two terms in equation (25) becomes zero, we can obtain the exact value of the product. Conversely, if we calculate the error terms for only a fixed number of iterations, we can obtain the approximate value of the product, sacrificing accuracy in return for reduction of computation time. The degree of accuracy of the result can thus be directly controlled by the number of iterations.

Figure 4 shows an implementation of the Iterative Logarithmic Multiplier. It contains two copies of most of the hardware intensive components in order to parallelized computation and reduce computational time, and as described in the next section, this is where the squaring unit gains its advantage.

## 5. SQUARING UNIT

In the previous section, we discussed how the Iterative Logarithmic Multiplier works, and an implementation for the same. When it is used for squaring a number rather than multiplying two numbers, the implementation becomes drastically simpler. From the definition of $N$ in equation (21), the square of $N$ can be represented as

$$N^2 = 4^k + 2^{k+1}(N - 2^k) + (N - 2^k)^2 \qquad (28)$$

When comparing equation (28) to (23), it becomes apparent how the representation of the square is much simpler than that of the product. First and foremost, instead of having two different values for $k$ and $x$, we just have one for each. That means that every operation that required two copies of the same hardware component (the priority encoder, the $k_1 + k_2$ bit adder, barrel shifter and the leading one detector) to parallelize computation now just requires one. That halves the hardware requirement for the biggest components of the multiplier. Also, no decoder is required since $4^k$ can be represented simply as $(100)_2 << k$.

The resulting architecture of the squaring unit is shown in Figure 5. Unlike in the case of the Iterative Logarithmic Multiplier, the adder and the barrel shifter units do not have to be used parallely, hence they can be reused in each stage further reducing the hardware complexity. As is evident, the hardware requirement for the squaring unit is less than half as compared to the basic multiplier unit of Figure 4.

## 6. POWERING UNIT

In the previous section, it was adequately emphasized that maximizing the use of a squaring unit will not only reduce the total hardware requirement, but also power consumption. Thus, the architecture for the powering unit was designed according to the heuristic "maximize squaring". It was pointed out in section I that every even power of a number $(x^{2k})$ is a square of some other lower power $(x^k)$. Hence, every even power is calculated by using a squaring unit, rather than a multiplication unit. Also, notice that every odd power of a number $(x^{k+1})$ can be represented as a product of the previous even power $x^k$ and the number $x$ itself. Since the priority encoder and Leading One Detector (LOD) values for $x$ are already calculated when calculating $x^2$, we can cache these values so that in every subsequent multiplication, the cached values are used. In that case, the multiplier unit would also require just one priority encoder and one LOD. Figure 6 shows a graphical representation of the above logic. To summarize

1) Calculate $x^2$ from $x$ using the squaring unit, and simultaneously cache the priority encoder values $(k)$ and the LOD values $(N - 2^k)$ for $x$.

2) In every cycle, repeat steps 3 – 5, until the desired precision is received.

3) Calculate the next odd power $x^{k+1}$ using the multiplier, setting its inputs as $x$ and $x^k$. For all calculations pertaining to $x$ (priority encoder and LOD), use the cached values.

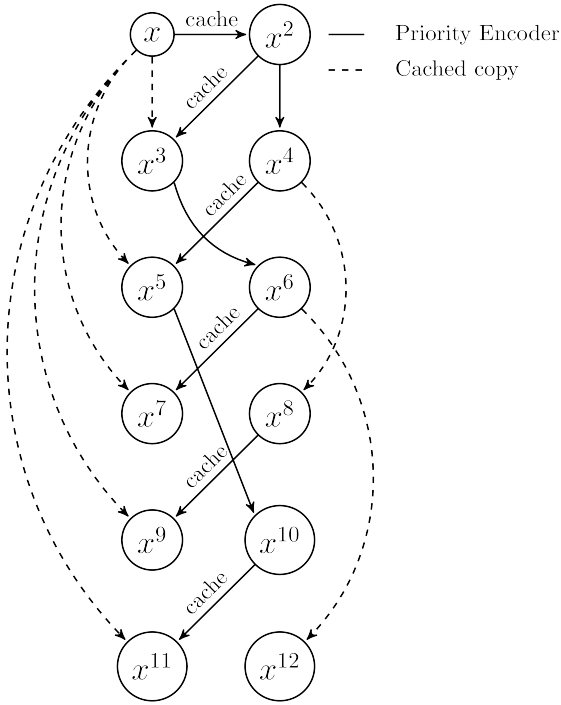 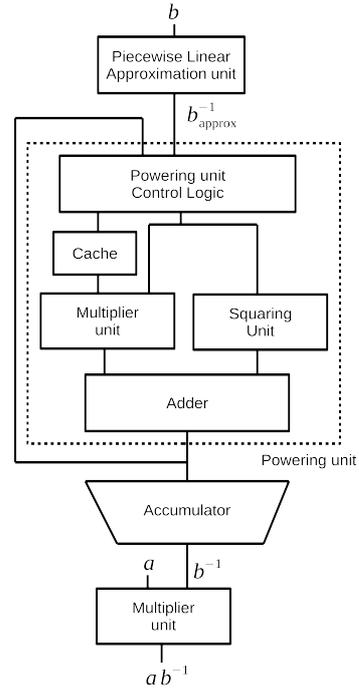

Figure 6: Flow diagram indicating the operation of the powering unit for calculating up to 12 powers of $x$.

Figure 7: System implementation

4) Calculate the next even power $x^{k+2}$ using the squaring unit, setting its inputs as $x^{(k+2)/2}$
5) If $(k+2)/2$ is even, use the cached priority encoder values.
6) Add the outputs from step 3 and step 4 to generate two iterations worth of correction in the approximation.

## 7. Conclusion

In this paper, we propose and investigate a new architecture for a floating point division unit. We show that the Taylor-series expansion algorithm can be used to generate approximations for the reciprocal of a number up to an arbitrary precision, and analyse the errors for the same. We propose a new piecewise linear approximation based method to generate the first approximation required by the Taylor-series expansion algorithm, and present an extensive analysis. We then present the architecture for a squaring unit derived from the Iterative Logarithmic Multiplier, and argue that it requires less than 50% hardware, as compared to the Iterative Logarithmic Multiplier. Finally, we present a cumulative implementation of the powering unit, and discuss some of the enhancements made in order to further boost its performance. The complete system is illustrated in Figure 7.

The performance of the system can be improved by pipelining the architecture for the Iterative Logarithmic Multiplier [12] and the squaring unit, but at the cost of increase in hardware utilization.